\documentclass[twocolumn]{aastex62}

\usepackage{graphicx}
\usepackage{subfigure}
\usepackage{booktabs}
\usepackage{amsmath}
\usepackage{amssymb}

\submitjournal{ApJ}

\shorttitle{diffusion of volatiles in ASW}
\shortauthors{He et al.}

\begin{document}

\title{Measurements of diffusion of volatiles in amorphous solid water: application to interstellar medium environments}

\correspondingauthor{Jiao He}
\email{jhe08@syr.edu}

\correspondingauthor{Gianfranco Vidali}
\email{gvidali@syr.edu}

\author[0000-0003-2382-083X]{Jiao He}
\affiliation{Physics Department, Syracuse University, Syracuse, NY 13244, USA}
\affiliation{Current address: Raymond and Beverly Sackler Laboratory for Astrophysics, Leiden Observatory, Leiden University, PO Box 9513, 2300 RA Leiden,The Netherlands}

\author{SM Emtiaz}
\affiliation{Physics Department, Syracuse University, Syracuse, NY 13244, USA}

\author{Gianfranco Vidali}
\affiliation{Physics Department, Syracuse University, Syracuse, NY 13244, USA}

\begin{abstract}
The diffusion of atoms and molecules in ices covering dust grains in dense clouds in interstellar space is an important but poorly characterized step in the formation of complex molecules in space. Here we report the measurement of diffusion of simple molecules in amorphous solid water (ASW), an analog of interstellar ices, which are amorphous and made mostly of water molecules. The new approach that we used relies on measuring in situ the change in band strength and position of mid-infrared features of OH dangling bonds as molecules move through pores and channels of ASW. We obtained the Arrhenius pre-exponents and activation energies for diffusion of CO, O$_2$, N$_2$, CH$_4$, and Ar in ASW. \added{The diffusion energy barrier of H$_2$ and D$_2$ were also measured, but only upper limits were obtained.} These values constitute the first comprehensive set of diffusion parameters of simple molecules on the pore surface of ASW, and can be used in simulations of the chemical evolution of ISM environments, thus replacing unsupported estimates. \added{We also present a set of argon temperature programmed desorption experiments to determine the desorption energy distribution of argon on non-porous ASW.}
\end{abstract}

\section{Introduction} \label{sec:intro}
In the Interstellar Medium (ISM), there are molecules that are complex enough to be considered as the building blocks of key molecules in the origin of life \citep{Chiar1997}. Astronomical observations, computer modeling, and laboratory experiments have all shown clearly that many of these interstellar complex organic molecules (ICOMs) are originated from the ice mantle covering cosmic dust grains in dense clouds \citep{Herbst2014}. In such environments where there is little penetration of UV light and the gas temperature is low ($<$50 K), gas-phase reactions are inefficient, and ices' advantage in storing molecules and radicals at low temperature ($\sim$10 K) for a long time, is clear. However, except for a few cases, the processes leading to the formation of these complex molecules in ices are either unknown or still insufficiently characterized. One poorly understood step in molecule formation is the diffusion of reactants in/on ices. \added{In the Langmuir-Hinshelwood mechanism, which is the most important mechanism in gas-grain astrochemical modeling, the rate of reactions is largely determined by the diffusion rate of reactants on the surface.} After gas phase radicals and molecules accrete on the ice mantle, they diffuse on the surface or penetrate into the ice to react with each other. The rate of diffusion governs how fast chemical reactions take place in the solid state and the abundance of ICOMs in the ice mantle. Yet, the process of diffusion under typical dense cloud conditions in the ISM is poorly characterized.

The most common molecule that has been detected in interstellar ices is water, followed by CO, CO$_2$, CH$_3$OH, CH$_4$, and NH$_3$. N$_2$ and O$_2$ should also be present, but \deleted{ being infrared inactive,} it is hard to establish their abundance \added{from the infrared} \citep{Boogert2015}. Observations of the absorption of the OH  vibrational stretch  in the infrared have shown that the structure of ice in the ISM is mostly amorphous \citep{Hagen1981}. In the laboratory, water ice grown by water vapor deposition at a temperature lower than $\sim$140 K is commonly referred to as amorphous solid water (ASW). Depending on the temperature, it may have a porous structure with a large surface area per unit volume, and the lower the deposition temperature is, the higher is the porosity \citep{Ayotte2001,Raut2007,Bossa2014}. To some extent, details of the amorphous ice morphology depend on the preparation methods \citep{Kimmel2001}. As the temperature of the ice is raised, pores collapse and the ice compacts \citep{Bossa2012}. In space, compaction can also be induced by cosmic rays \citep{Palumbo2005,Raut2008}. The non-detection in the ISM of dangling OH (dOH) bands in the infrared \citep{Keane2001} ---indicative of porous ASW \citep{Buch1991}---seems at first to argue against the porous hypothesis. However, experiments by  \citet{Bossa2014} found that the missing of dOH bonds does not necessarily mean that the ice is fully compact. The presence of these pores in interstellar ices is important because they have the potential to absorb a large \replaced{amount}{number} of molecules, and are also the conduits for molecules to move and undergo chemical reactions. The very rich chemistry in the interstellar space seems to agree with the hypothesis that the ice mantle is porous, at least to some extent.  Diffusion of molecules or radicals on the pore surface of interstellar ices is a key process in order to understand chemical reactions in the solid state in the ISM\@. In this work, we study the diffusion of simple molecules in ASW with the purpose to obtain parameters of the diffusion process, such as the pre-exponent factor $D_0$ and activation energy $E_{\rm dif}$ in the Arrhenius expression for the diffusion coefficient $D(T)=D_0 e^{-\frac{E_{\rm dif}}{kT}}$, that are important in simulations of the chemical evolution of ices \citep{Garrod2006, Garrod2008, Garrod2011}.

There exist a few attempts to quantify the diffusion of simple molecules on/in ices. \citet{Livingston2002} used laser resonant desorption to follow the bulk diffusion of NH$_3$ and CH$_3$OH molecules from a layer of NH$_3$ or CH$_3$OH sandwiched in water ice between 140 and 180 K. This temperature region is much higher than the typical temperature in dense clouds, and it is hard to extrapolate their experimental data to dense cloud conditions. \citet{Mispelaer2013} deposited a thick layer of water on top of target molecules CO, NH$_3$, H$_2$CO, or HNCO. By heating, the target molecules migrated through the ice and desorbed into vacuum. Infrared absorption of these molecules was measured in order to quantify the desorption of molecules from the ice. Fick's law of diffusion was used to simulate the experiments, and the diffusion energy barriers as well as the pre-exponential factors in the diffusion \replaced{rate}{coefficient} were obtained. Their work is an important step toward the understanding of diffusion in water ice. However, it suffers from two complications: 1) In their experiments, desorption and diffusion are intimately entwined, and it is hard to separate the effect of one from the other; 2) During the measurement, the water ice structure may be changing; it is unclear how the changes affect the experiments. Following a similar procedure, \citet{Karssemeijer2014} measured the diffusion of CO in porous water ice. \citet{Lauck2015} extended the work of \citet{Karssemeijer2014} and presented a systematic study of CO diffusion in ASW. Rather than looking into the desorption of CO from the ice as in \citet{Mispelaer2013}, \citet{Lauck2015} distinguished the IR absorption of CO molecules that  interacted with CO from those interacting with the surface of pore in water ice. By performing a set of isothermal experiments, they quantified the diffusion of CO in porous ASW\@. They attributed the diffusion to be surface diffusion along pores.  \citet{Lauck2015} avoided the first complication of \citet{Mispelaer2013} mentioned above, but not the second one. Most recently, \citet{Cooke2018} utilized the change in the CO$_2$ $\nu_3$ band profile induced by CO as a probe to study the diffusion of CO in CO$_2$ ice. They found the diffusion energy barrier to be $300\pm40$ K, which is 0.18--0.24 times of the binding energy of CO \added{on non-porous ASW at low surface coverages} \citep{He2016}. In all of the three studies, the mechanism was found to be diffusion along the pore surfaces instead of diffusion in bulk ice. \added{The penetration of molecules through porous media is also called percolation. In this work, we adopt the term ``diffusion''  that is more familiar to the astronomical community.}

Rigorously speaking, the above experiments do not necessarily measure the same diffusion processes as in interstellar conditions. In a typically experiment to measure diffusion, the adsorbate-adsorbate interaction dominates over adsorbate-adsorbent interaction, and the diffusion is mostly concentration driven. In contrast, under realistic interstellar conditions, the surface coverage of adsorbate is usually very low, and the adsorbate-adsorbate interaction is negligible. In this case, one would like to measure the tracer diffusion coefficient, which describes the random walk of a particle. To approximate this condition and obtain diffusion parameters that are applicable to interstellar conditions,  the surface coverage of that particle  in an experiment has to be low, unless the adsorbate-adsorbate interaction is significantly weaker than the adsorbate-adsorbent one. In a prior work, we studied the diffusion of low coverage CO$_2$ on compact ASW surface \citep{He2017}. Although the CO$_2$-CO$_2$ interaction was found to be stronger than the CO$_2$-H$_2$O interaction, in the experiment the CO$_2$ coverage was limited to 0.23 monolayer (ML). The diffusion parameters should still apply to interstellar conditions. However, due to limited experimental data, we couldn't determine the pre-exponent factor $\nu$ and the diffusion energy barrier $E_{\rm dif}$ in the diffusion rate $\Gamma=\nu e^{-\frac{E_{\rm dif}}{kT}}$ at the same time. The diffusion energy barrier $E_{\rm dif}$ was calculated assuming the pre-exponential factor $\nu$ of 10$^{12}$ s$^{-1}$, which is the typical value of the vibrational frequency of a weakly adsorbed particle on a surface. In the study reported here,  we measure the diffusion rate $D(T)$ of CO, N$_2$, O$_2$, CH$_4$, and Ar on the surface of pores in porous ASW, and determine both the pre-exponent factor $D_0$ and activation energy barrier $E_{\rm dif}$ for the  diffusion. For these adsorbate molecules, the adsorbate-adsorbent interaction is about 1.5--2 times stronger than the  adsorbate-adsorbate interaction \citep{He2016}. We also chose the deposition dose so that the coverage is less than full coverage of the surface area of the ASW sample, as an effort to study the diffusion that is close to that in the interstellar condition.

In the next Section, the experimental results are presented, followed by modeling of the experimental  using Fick's law of diffusion. Finally, we comment on how these results should be used in simulations of the chemical evolution of ISM ices.

\section{Experimental}
Experiments were performed using a ultra-high vacuum (UHV) setup located \replaced{in}{at} Syracuse University. \added{A} detailed description can be found in prior works \citep{He2018a, He2018b}; here we briefly summarize the features that are relevant to this study. \added{The main ultra-high vacuum (UHV) chamber reaches a base pressure of $4\times10^{-10}$ torr.} At the center of the UHV chamber, a gold coated copper disk was used as the sample onto which ice was grown. The sample can be cooled down to 4.8~K using a closed cycle helium cryostat. A cartridge heater installed right underneath the sample is used to heat the sample. A silicone diode \added{located right underneath the sample disk and} a Lakeshore 336 temperature controller was used to measure the temperature to an accuracy better than 50 mK and to control the heating of the sample. Water vapor and other gas molecules were deposited from two separate UHV precision leak valves. Distilled water underwent at least three freeze-pump-thaw cycles before being sent into the chamber. The gas/vapor deposition was controlled accurately by using a stepper motor connected to the leak valves. A LabVIEW program measures the chamber pressure and calculates the thickness in real time. Typically, the relative accuracy of gas deposition is better than 0.1\%. Water deposition has a larger uncertainty of 1\% because of the instability of the water inlet pressure. For all of the experiments in this study, gas was deposited from the background. \added{The deposition dose was calculated from the impingement rate ($IPR$):
\begin{equation}
IPR=\frac{P}{\sqrt{2\pi m k_{\rm B} T}}
\end{equation}
where P is the chamber pressure, $m$ is the mass of gas molecule, and $T$ is the gas temperature (assumed to be room temperature). The impingement rate has the unit of molecules per unit surface area per unit time.} It can be converted to ML/s by assuming 1 ML = 10$^{15}$ cm$^{-2}$. The chamber pressure was recorded using a hot cathode ion gauge. For each gas, the measured pressure was corrected for the ionization cross section of that gas. The main source of uncertainty in thickness comes from the pressure gauge. An absolute uncertainty of up to 30\% is possible for this type of gauge. A Nicolet 6700 FTIR in the Reflection Absorption Infrared Spectroscopy (RAIRS) setup was used to monitor the ice. The FTIR scans and averages 8 spectra every 10 seconds at a resolution of 0.5 or 1.0 cm$^{-1}$.

A set of volatile gases and water vapor co-deposition experiments were performed to find out how the inclusion of volatile molecules in the water ice affects the dOH bands of water. 100 ML of water and 30 ML of volatile molecules were deposited simultaneously onto the gold sample when  the surface was at 10~K. \added{The RAIR spectra were measured after deposition without warming up the ice. } The volatile molecules used in this study are: CO, CH$_4$, N$_2$, O$_2$, Ar, D$_2$, and  H$_2$.

The shifting of dOH by a volatile molecule was exploited to measure the diffusion  on the ASW surface. Water was deposited from the background when the sample was at 10 K. After 200 ML of water deposition, the ice was annealed at 70 K for 30 minutes before further experiments.  The purpose of annealing the ice to 70 K is:
\begin{itemize}
  \item To ensure the water ice structure does not change during the whole diffusion experiment. \added{We tried to anneal an ASW sample at 70 K for 2.5 hours, and found that most of the changes in dOH happens during the first 30 minutes. When the ice is cooled down to 10--25 K range for diffusion measurement, the change in ice structure should be negligible.} Our measurements of diffusion are different from the experiments by \citet{Lauck2015}, in which CO diffusion and ASW pore collapse happen simultaneously, and it was not possible to separate the effect of pore collapse from the diffusion of CO\@.
  \item After annealing at 70 K, the 2-coordinated dOH band at 3720 cm$^{-1}$ is almost gone, and only the 3-coordinated dOH band persists; we only need to deal with the shifting in position of one band instead of two bands, therefore making the data analysis simpler. More discussion about the shifting of dOH are in the next section.
  \item At 70 K, the pore surface area as well as the dOH band intensity are still appreciable, giving reasonable signal to noise ratio for the diffusion measurement. This is the reason to anneal at 70 K instead of at higher temperatures.
\end{itemize}
After annealing at 70 K, the ASW samples were cooled down to 10 K for the deposition of a volatile gas, except for H$_2$ and D$_2$, which were deposited when the annealed ASW was at 4.8 K. The dOH band of the annealed water is red shifted from 3697 cm$^{-1}$ to $\sim$3694 cm$^{-1}$ after annealing. These volatile molecules deposited on top of the annealed water ice are not mobile (except for H$_2$ and D$_2$), and they build up as pure layers of ice. \added{Just after the deposition of the volatiles} the dOH absorption is mostly at the pure water dOH position ---3694 cm$^{-1}$, with a small fraction of shifted dOH due to the penetration of molecules in the topmost layers. The  dose of CO, N$_2$, and Ar were fixed at 16 ML, and those for  CH$_4$ and O$_2$  at 20 ML. According to our measurement \citep{He2018b}, 20 ML of molecules occupy most of the dOH sites on the pore surface. \added{After deposition at 10 K, we waited until the residual volatile molecules in the chamber were completely pumped away before} heating up the ice quickly at a ramp rate of 0.5 K/s to the desired isothermal experiment temperature and remained at this temperature for 30 minutes. The overshooting in temperature is less than 0.2~K. RAIRS spectra were collected in a 10 seconds time resolution to monitor the change in dOH bands.

\section{Results}
\subsection{Co-deposition of volatiles and water}
\begin{figure}[tbp]
 \centering
 \includegraphics[width=1\columnwidth]{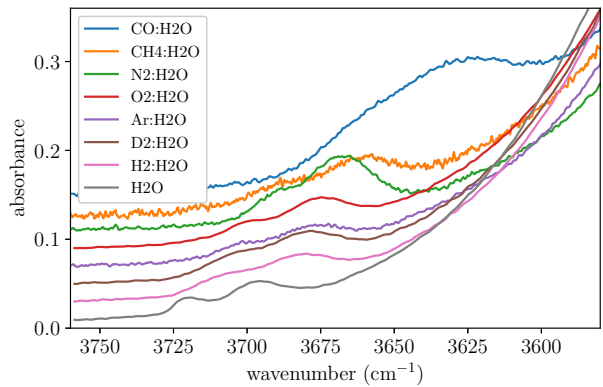}
 \caption{Reflection Absorption InfraRed Spectra (RAIRS) of the dOH region for pure water, and the mixture of water and other molecules. \added{All ices are deposited from the background when the surface is at 10 K, and measured at the same temperature. } The spectra are normalized to the maximum of the bulk water OH stretch peak and then offset for clarity. }
 \label{fig:co-dep}
\end{figure}

The infrared dOH absorption spectra of water and volatile molecule mixtures are shown in Figure~\ref{fig:co-dep}. The spectrum of pure water is also shown for comparison. The dOH bands of pure water are located at $\sim3696$ cm$^{-1}$ and $\sim3720$ cm$^{-1}$, for 3- and 2-coordinated dOH bonds \citep{Buch1991}, respectively. When water ice is mixed with volatile molecules, a red shift of both dOH bands can be seen. For CO and water ice mixture, the shifted dOH band is a broad peak centered at 3635 cm$^{-1}$. It is possible that this broad peak consists of two components close to each other.  The shifted dOH peaks are fitted with two Gaussian functions (except for CO). Their positions are listed in Table~\ref{tab:doh}.

\begin{table}[tbp]
\centering
\caption{Peak positions of ASW dOH bands and the shifted positions in the mixture of water with a volatile molecule. \added{All ices are deposited on a substrate at 10 K, and measured at the same temperature.}}
\label{tab:doh}
\begin{tabular}{lr}
    &   dOH band positions / cm$^{-1}$     \\ \midrule
H$_2$O         & 3697, 3720 \\
CO:H$_2$O      &  3635 \\
N$_2$:H$_2$O   & 3668, 3692 \\
O$_2$:H$_2$O   & 3676,  3701 \\
Ar:H$_2$O      & 3679,  3702 \\
CH$_4$:H$_2$O  & 3661, 3688  \\
D$_2$:H$_2$O   & 3681, 3705 \\
H$_2$:H$_2$O   & 3684, 3708
\end{tabular}
\end{table}

\subsection{Diffusion of volatile molecules on annealed water ice}
\begin{figure}[tbp]
 \centering
 \includegraphics[width=1\columnwidth]{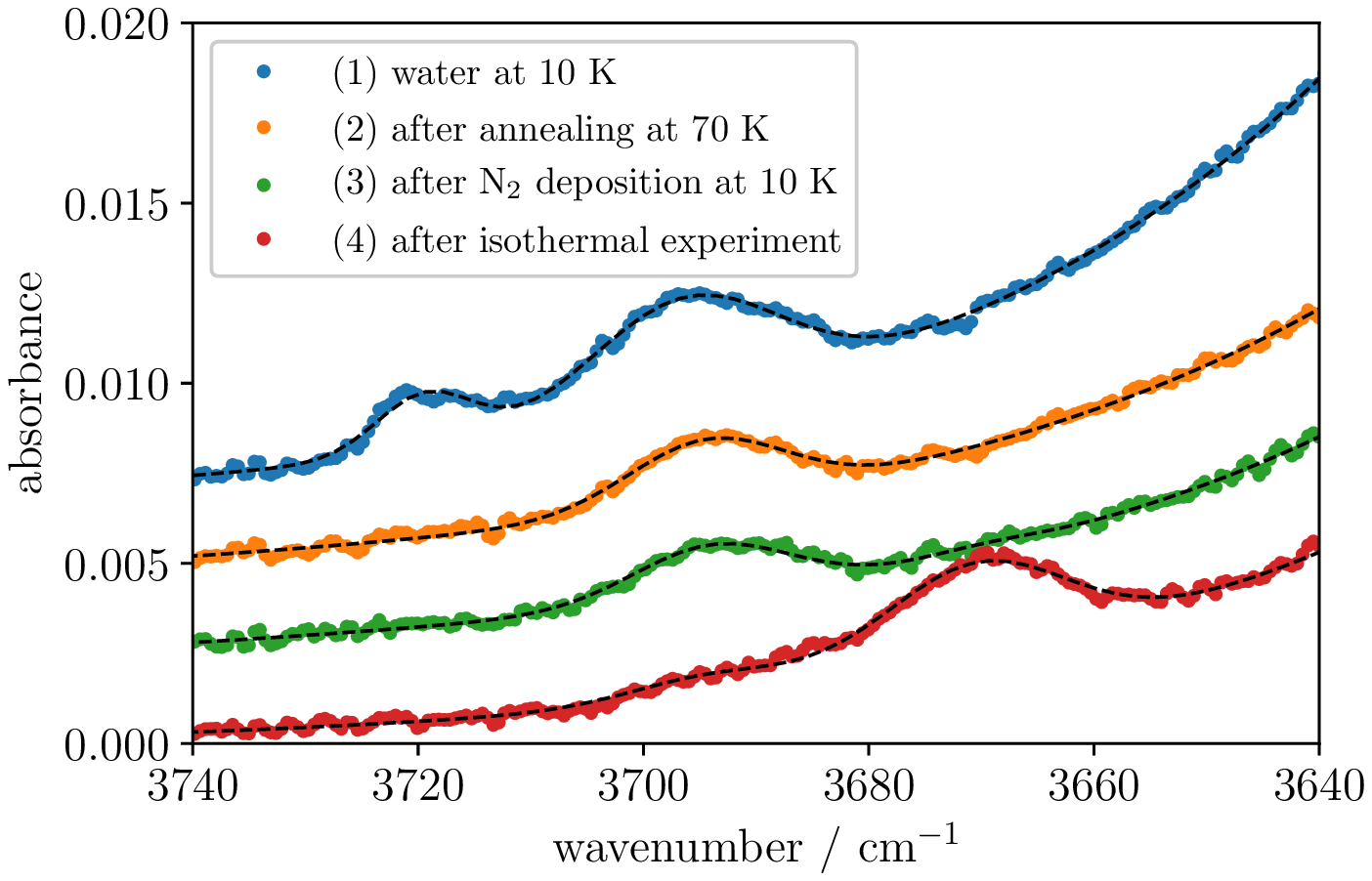}
 \caption{\replaced{Example of fitting the spectrum of N$_2$ on ASW before and after annealing. The dashed lines is the fit while the solid circles are the experimental data. }{RAIR Spectra of: (1) 200 ML pure water ice deposited at 10 K; (2) after annealing at 70 K for 30 minutes; (3) after cooled down to 10 K and then deposited 16 ML N$_2$ on the top; (4) after isothermal experiment. The dashed lines are the fit while the solid circles are the experimental data. Spectra are offset for clarity. }}
\label{fig:n2_fit_example}
\end{figure}

\begin{figure}[tbp]
 \centering
 \includegraphics[width=1\columnwidth]{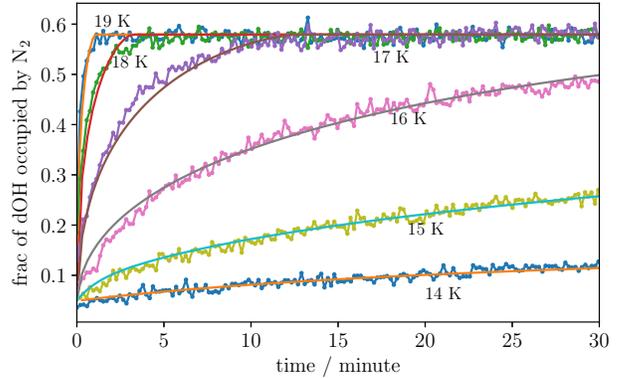}
 \caption{The band area of the 3672 cm$^{-1}$ peak during \replaced{annealing}{isothermal} experiments of N$_2$ on ASW at different \added{target} temperatures. The lines are fits using the model described in the Section~\ref{sec:model}. }
\label{fig:n2_ann_area}
\end{figure}

\begin{figure}[tbh]
 \centering
 \includegraphics[width=1\columnwidth]{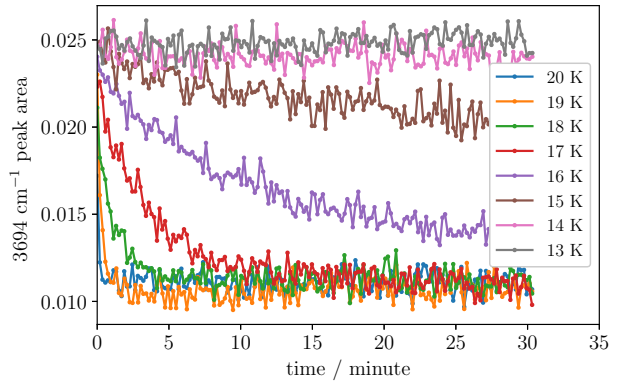}
 \caption{The band area of the 3694 cm$^{-1}$ peak during the \replaced{annealing}{isothermal} experiment of N$_2$ on ASW at different \added{target} temperatures. }
 \label{fig:n2_ann_3694_area}
\end{figure}

Figure~\ref{fig:n2_fit_example} shows examples of the dOH band during the N$_2$ diffusion experiments. \added{After water deposition at 10 K, both dOH bands at 3697 cm$^{-1}$ and 3720 cm$^{-1}$ are present. After  annealing the ASW at 70 K for 30 minutes, the dOH band at 3720 cm$^{-1}$ is mostly gone, and the other one redshifts slightly from 3697 cm$^{-1}$ to 3694 cm$^{-1}$. Then ASW was cooled down to 10 K for N$_2$ deposition.} The N$_2$ molecules are mostly on top of the ASW in the pure form. Only a small fraction of N$_2$ are interacting with the water ice surface. After \replaced{annealing}{isothermal} experiment at a sufficiently high temperature \added{(but still lower than the desorption temperature)} for a set period of time, N$_2$ diffuses into the ASW along the pore surfaces, and evenly covers the pore surface of the ASW. The original dOH band at 3694 cm$^{-1}$ is mostly shifted to 3672 cm$^{-1}$, which is blue shifted by a few cm$^{-1}$ with respect to \added{the shifted 3-coordinated dOH at} 3668 cm$^{-1}$ as shown in Table~\ref{tab:doh}. Other molecules also show similar blue shifts as N$_2$, i.e., the shifted dOH positions on annealed ASW are slightly shifted respect to the \added{shifted 3-coordinated } dOH positions in Table~\ref{tab:doh}. Figure~\ref{fig:n2_fit_example} also shows the fitting of the spectra. A broad Gaussian and a broad Lorentzian function are used to fit the blue wing of the bulk water absorption. The original dOH band and the shifted one are each fit using a Gaussian function. This fitting scheme is applied to each spectrum collected in the whole experiment to trace the diffusion of N$_2$ in the ASW. The band area of the shifted dOH band at 3672 cm $^{-1}$ and the original dOH band at 3694 cm$^{-1}$ during the isothermal experiment at each target temperature are shown in Figure~\ref{fig:n2_ann_area} and Figure~\ref{fig:n2_ann_3694_area}, respectively. It is evident that the shifting of dOH strongly depends on the target temperature.

\section{Modeling}
\label{sec:model}
\begin{figure}[tbp]
 \centering
 \includegraphics[width=1\columnwidth]{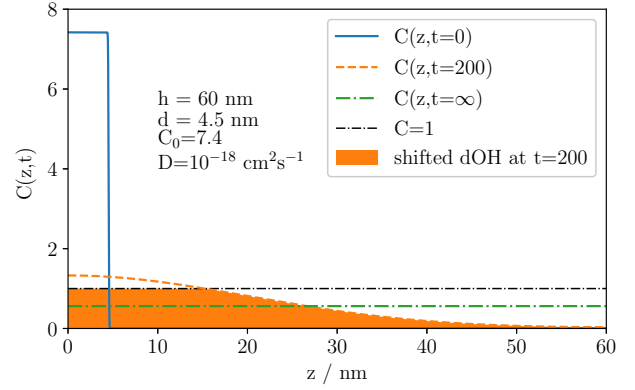}
 \caption{Simulations based on Eq~\ref{eq:solution}. \added{The simulated $C(z,t)$ distributions for $t$=0 s, 200 s, and $\infty$ s are shown. }The shaded area is proportional to the shifted dOH band area at $t=200$ s. }
\label{fig:fick_sim}
\end{figure}

Fick's law of diffusion was used to obtain the diffusion energy barrier as well as the pre-exponential factor $D_0$, similar to  other studies \citep{Mispelaer2013,Karssemeijer2014,Lauck2015}. The one dimensional Fick's law of diffusion is described by the following equation:
\begin{equation}
  \frac{\partial C(z,t)}{\partial t}=D(T) \frac{\partial^2 C(z,t)}{\partial^2 z} \label{eq:fick}
\end{equation}
where $C(z,t)$ is the coverage \added{(unitless)} at depth $z$ into the ice. \added{If the amount of of N$_2$ at depth $z$ is just enough to cover the whole pore surface at that depth, then $C(z,t)=1$.} $D(T)$ is the temperature dependent diffusion coefficient. We assume that 1 ML of water corresponds to a thickness of 0.3 nm. Therefore the total water ice thickness is $h=60$ nm. Right after N$_2$ deposition at 10 K, N$_2$ should only cover the top  few layers. We denote this initial penetration depth as $d$. After sufficient diffusion, N$_2$ penetrates into the ASW and evenly covers the pore surfaces. We have the following equations:
\begin{equation}
  C(z,t=0) = \begin{cases}
C_0 & 0<z<d \\
0   & d<z<h
  \end{cases}  \\
\end{equation}
\begin{eqnarray}
  C(z,t=\infty) = C_{inf}\\
  C_0 d = C_{inf}h
\end{eqnarray}
\begin{figure}[tbh]
 \centering
 \includegraphics[width=1\columnwidth]{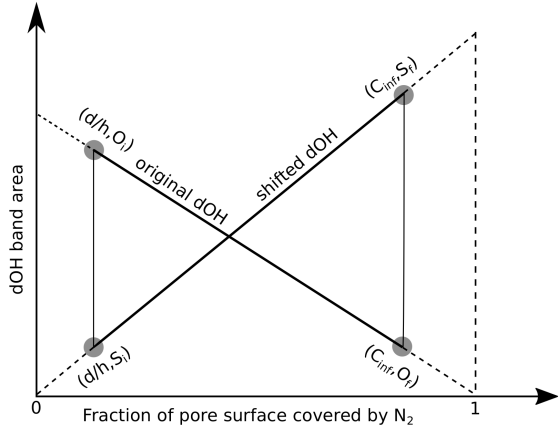}
 \caption{Illustration of the change of  the area of the dOH bands. C$_{inf}$ is the coverage at t=${inf}$ and O and S are the unshifted and shifted dOH bands, see text.}
\label{fig:drawing}
\end{figure}
The values of $d$, $C_0$ and $C_{inf}$ can be obtained from Figure~\ref{fig:n2_ann_area} and Figure~\ref{fig:n2_ann_3694_area}. We denote the area of the original dOH band at 3694 cm$^{-1}$ before and after isothermal experiment (such that N$_2$ covers the pore surface evenly) to be $O_i$ and $O_f$, respectively. Similarly, we denote the area of the shifted dOH band at 3672 cm$^{-1}$ before and after sufficient diffusion to be $S_i$ and $S_f$, respectively. Figure~\ref{fig:drawing} illustrates how the two dOH bands change with the fraction of pore surface covered by N$_2$. Simple calculations show that:
\begin{eqnarray}
C_{inf} = \frac{S_f(O_i-O_f)}{O_i S_f-O_f S_i}\\
\frac{d}{h}=\frac{S_i(O_i-O_f)}{O_i S_f-O_f S_i}
\end{eqnarray}
In the model we don't consider desorption, because in all of the isothermal experiment, the ice temperature is always below the temperature at which desorption occurs. Solving Eq~\ref{eq:fick} yields:
\begin{equation}
  C(z,t)=C_{\inf}+\sum\limits_{n=1}^{\infty}\frac{2h}{n\pi d} \sin(\frac{n\pi d}{h})\times \cos(\frac{n\pi z}{h})\exp(-\frac{n^2 \pi^2}{h^2} Dt)  \label{eq:solution}
\end{equation}
Figure~\ref{fig:fick_sim} shows the solution at $t$=0, 200, and $\infty$ s. At $t=0$ s, N$_2$ molecules  penetrate only to depth $d$. The shifted dOH band area should be proportional to $d/h$. In this specific case, the coverage is $C(z<d,t=0)=7.4$.  At the other extreme, $t=\infty$, N$_2$ molecules are evenly distributed on the pores. The shifted dOH band area is proportional to $C_{inf}$. At an intermediate stage $t=200$ s, the shifted dOH band area is proportional to the shaded area under $C=1$ and $C(z,t=200)$. Figure~\ref{fig:n2_ann_area} shows the fitting of the curves in Figure~\ref{fig:n2_ann_area}.

\begin{figure*}
\gridline{\fig{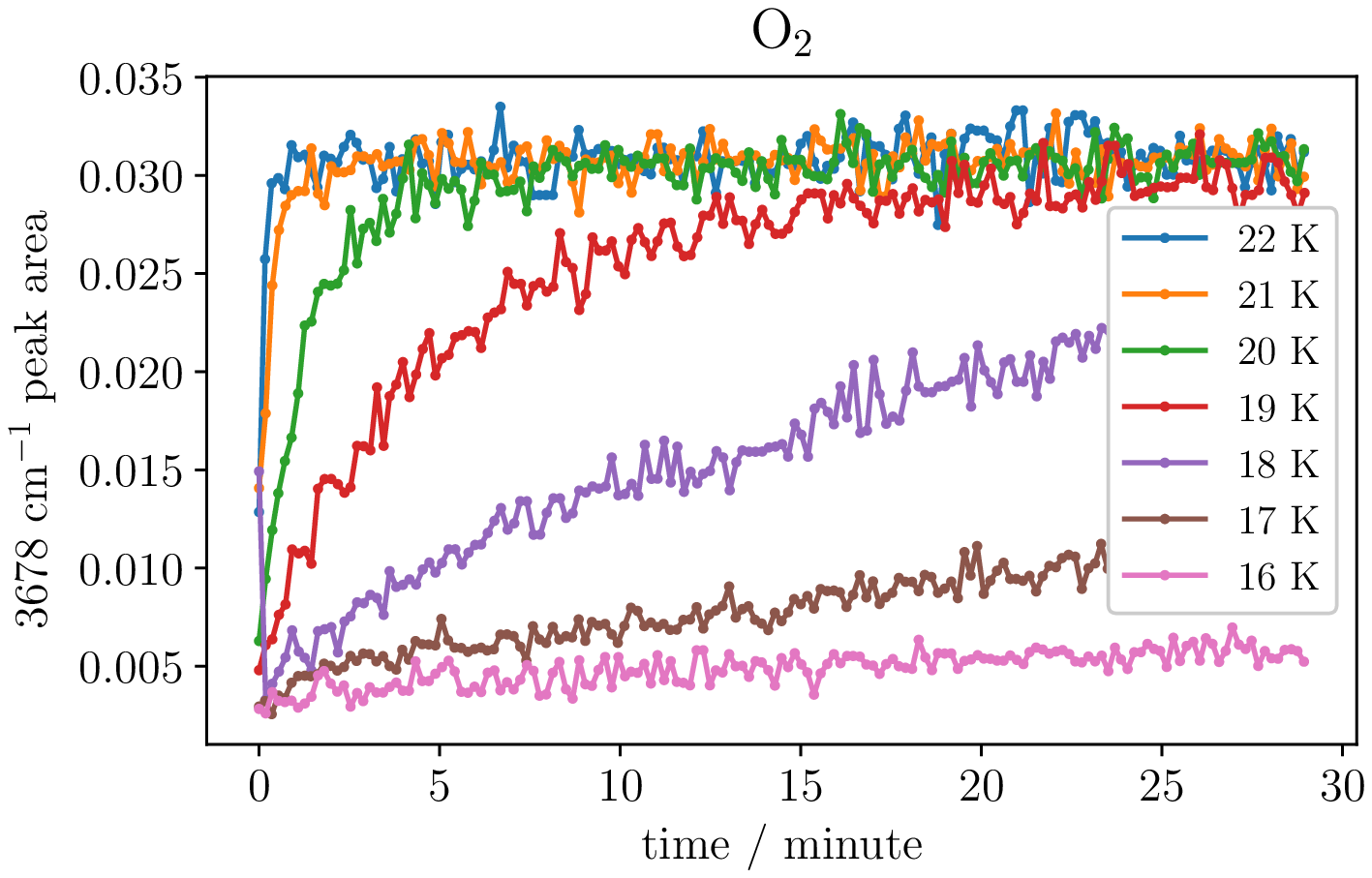}{0.5\textwidth}{}
          \fig{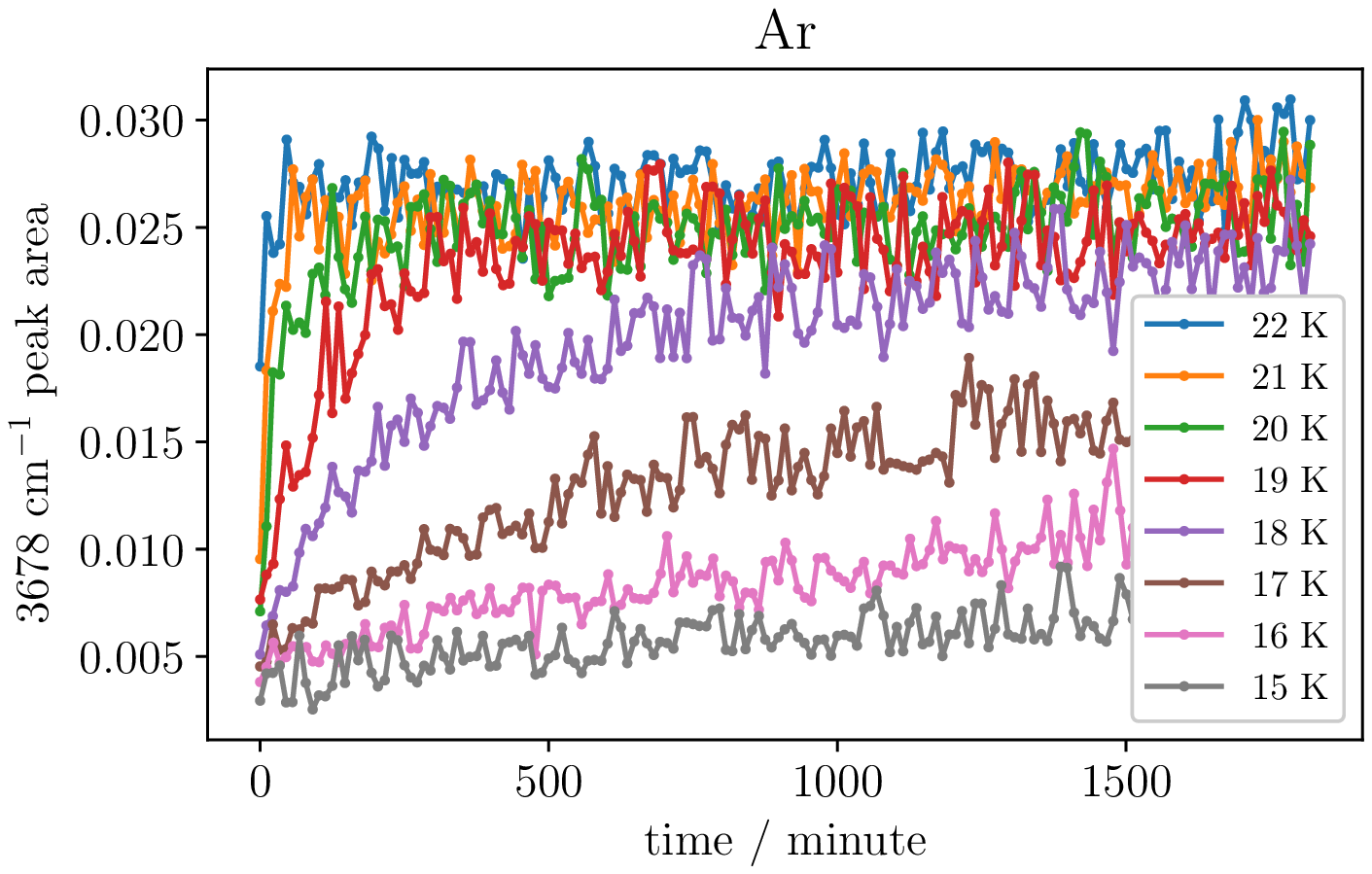}{0.5\textwidth}{}
          }

\gridline{\fig{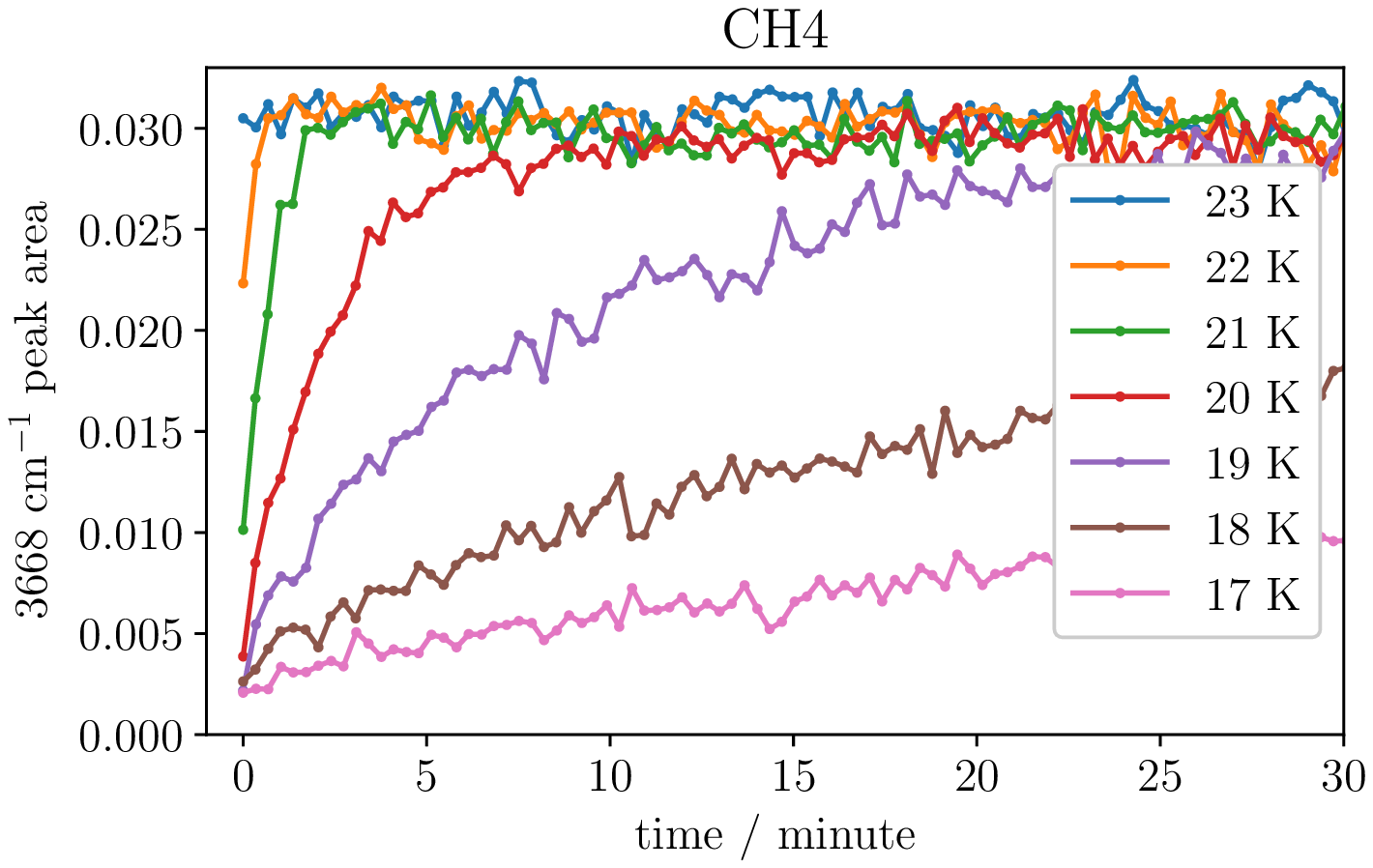}{0.5\textwidth}{}
          \fig{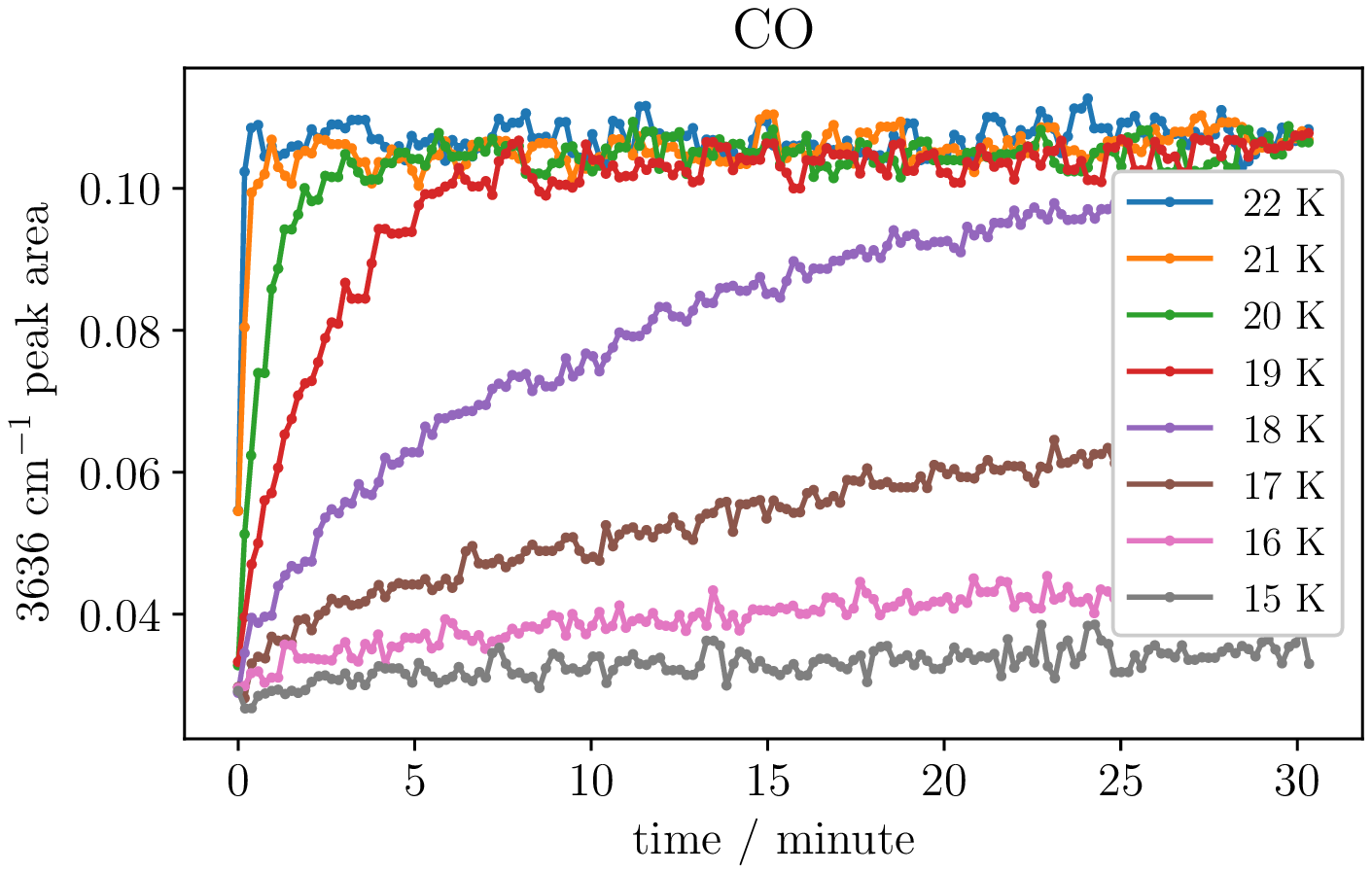}{0.5\textwidth}{}
          }
	  \caption{The band area of the shifted dOH peak during \replaced{annealing}{isothermal} experiments of O$_2$, Ar, CH$_4$ and CO on ASW at different \added{target} temperatures.}
\label{fig:other_mols}%
\end{figure*}

\begin{figure}[tbh]
 \centering
 \includegraphics[width=1\columnwidth]{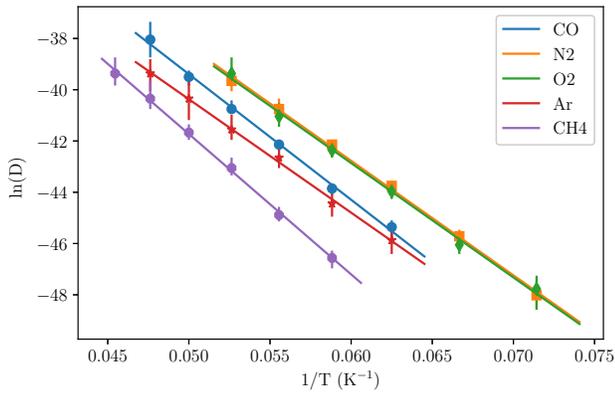}
 \caption{Arrhenius-type plot of the diffusion \replaced{rate}{coefficient} over 1/T and linear fittings. }
\label{fig:arr}
\end{figure}

Similar fitting procedures were applied to O$_2$, Ar, CH$_4$, and CO\@. The shifted dOH band area versus \replaced{annealing}{isothermal experiment} time at different \replaced{annealing}{target} temperatures is shown in Figures~\ref{fig:other_mols}.
After obtaining the $D(T)$ values from the fitting, we use the Arrhenius expression $D(T)=D_0 \exp(-E_{\rm dif}/T)$ to find  the pre-exponential factor $D_0$ and diffusion energy barrier $E_{\rm dif}$. In a $\ln(D_0)$ vs $1/T$ plot, see Figure~\ref{fig:arr},  data are fitted very well with the Arrhenius expression.  The best fitting diffusion energies and pre-exponents  are summarized in Table~\ref{tab:arr}.

We also attempted to measure the diffusion energy barrier of H$_2$ and D$_2$ on ASW\@. However, the diffusion rate of these two molecules are very high even at 4.8~K. They penetrate into the pores during the deposition stage. We could not establish the \added{exact} diffusion rate \added{of them} by isothermal \replaced{annealing}{experiment}. \added{Here we try to estimate the upper limit of their diffusion barriers. Figure~\ref{fig:arr} and Table~\ref{tab:arr} show that the values of prefactor $D_0$ are surprisingly similar for the molecules that are measured. We assume that the $D_0$ values for D$_2$ and H$_2$ are both 10$^{-7}$ cm$^{2}$s$^{-1})$. The diffusion of them is so fast that we could not see any time lag in shifting of dOH. Therefore the diffusion coefficient should be at least comparable with the higher values of $D$ shown in Figure~\ref{fig:arr}. We take the value $ln(D)>-41$. Using the formula $E_{\rm dif}=-Tln(D/D_0)$, we can estimate that $E_{\rm dif}<120$ K for both H$_2$ and D$_2$. These upper limits of H$_2$ and D$_2$ should be useful in astrochemical modeling. }

\begin{table}[tbh]
\centering
\caption{Pre-exponent \added{D$_0$} of the diffusion coefficient \replaced{$D_0$}{$D$} and diffusion energy barriers for the diffusion of molecules on the surface of ASW obtained from this study.}\label{tab:arr}
\begin{tabular}{lcl}
    & log$_{10}$(D$_0$/cm$^{2}$s$^{-1})$  & $E_{\rm dif}$ / K     \\ \midrule
CO  & -6.47$\pm$0.29  & 490$\pm$12 \\
N$_2$  & -6.93$\pm$0.25  & 447$\pm$9  \\
O$_2$  & -6.99$\pm$0.33  & 446$\pm$12 \\
Ar  & -7.92$\pm$0.24  & 443$\pm$10 \\
CH$_4$ & -6.23$\pm$0.23  & 547$\pm$10
\end{tabular}
\end{table}

\section{Discussion and Conclusion}
\begin{figure}[tbp]
 \centering
 \includegraphics[width=1\columnwidth]{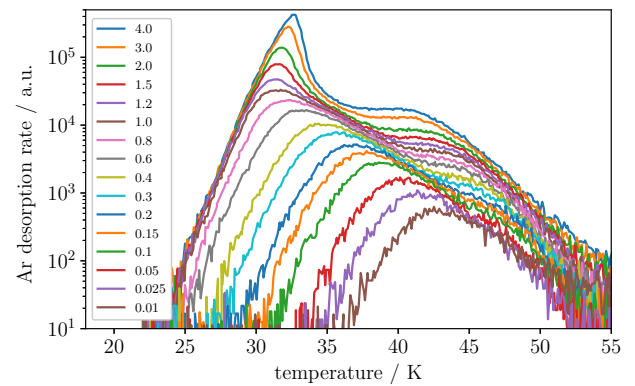}
 \caption{\added{Temperature programmed desorption (TPD) spectra of Ar from non-porous amorphous solid water (np-ASW). The ramp rate during TPD is 0.6 K$\cdot$ s$^{-1}$. The coverage in ML is shown in the inset. }}
\label{fig:ar_tpd}
\end{figure}

\begin{figure}[tbp]
 \centering
 \includegraphics[width=1\columnwidth]{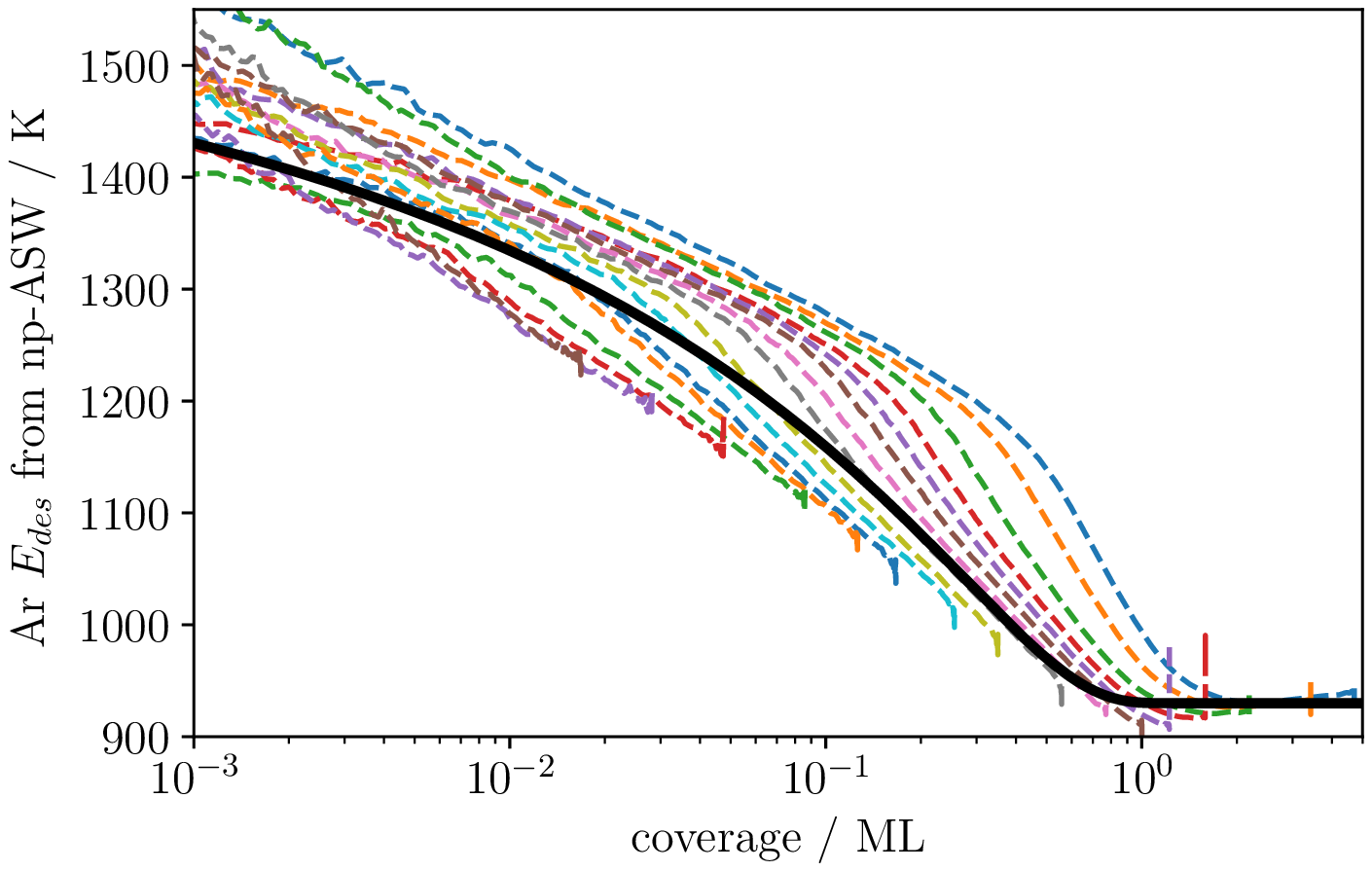}
 \caption{\added{The desorption energy distribution of Ar on np-ASW obtained from direct inversion of the TPD curves in Figure~\ref{fig:ar_tpd}. The direct inversion procedure is described in \citet{He2016}. }}
\label{fig:ar_Edes}
\end{figure}

The pre-exponential factor and energy barrier \replaced{for}{of} diffusion \added{coefficient} are key parameters in simulations of chemical evolution of ISM environments.  Because of the well-known difficulty, very few laboratory measurements of them are available in the literature. The values that are appropriate for tracer diffusion are even more scarce. Astrochemical models usually assume that the pre-exponent \added{$\nu$} of the diffusion rate \added{$\Gamma$} is the same as that for desorption, and that the $E_{\rm dif}$/$E_{\rm des}$ ratio is  \replaced{a universal constant} {the same for all volatiles}. This ratio \replaced{was}{is} often taken to be between 0.3 and 0.5. In Table~\ref{tab:ratios} we listed the diffusion energy barriers obtained from this study and desorption energies obtained \added{previously} by our group \deleted{previously} \citep{He2016}, except for Ar. Following the same procedure as in \citet{He2016}, we did a set of Ar temperature programmed desorption (TPD) experiments on non-porous ASW to obtain the desorption energy distribution of Ar. \added{The TPD spectra and resulting desorption energy distribution are shown in Figure~\ref{fig:ar_tpd} and Figure~\ref{fig:ar_Edes}, respectively. At relative high Ar coverages, there is a second desorption peak between 40 and 45 K. This component is perhaps due to the desorption from sample holder. In fitting the binding energy distribution using a empirical formula, we intentionally placed a higher weight for the TPD curves corresponding to lower coverages. We found that the desorption energy distribution can be fit by the formula:
\begin{equation}
E_{\mathrm{b}}(\theta)= 930 + 800 \exp \left(-\frac{1.5}{\max(0.2-\lg (\theta), 0.001)}\right)
\label{eq:ar_Edes_fit}
\end{equation}
where $\theta$ is the coverage. }

The binding energy for all of the volatile molecules covers a wide range. At low coverages, molecules preferably occupies deep adsorption sites, and the binding energy is high. As coverage increases, deeper sites are no longer available, and molecules \replaced{occupies}{occupy} shallower sites. At 1 ML and above, the binding energy is the lowest. The lower value of binding energy in Table~\ref{tab:ratios} corresponds to the value for 1 ML, and the higher values corresponds to very low coverages. The $E_{\rm dif}$/$E_{\rm des}$ ratios are also shown in the table. The ratio is mostly between 0.3 and 0.6, which agrees with \added{prior off-lattice kinetic Monte Carlo simulations by \citet{Karssemeijer2014diffusion} and} the range that has been assumed --- based solely on empirical estimates --- in models of the chemical evolution of ISM environments \citep{Garrod2011}. In our previous work \citet{He2017}, we measured the diffusion of CO$_2$ on non-porous ASW and found that the diffusion energy barrier is about 0.95 times the binding energy. It still remains a question why CO$_2$ is very different from the more volatile molecules. One possibility is that in \citet{He2017} the diffusion pre-exponent factor $\nu$ in the diffusion rate was assumed to be the same as the vibration frequency \added{$\nu_0$} of a single particle in the potential well of the surface, which is probably not true, since it ignores entropic effects and interaction with other particles \citep{Ehrlich1994}. The pre-exponent $D_0$ of the diffusion coefficient in Table~\ref{tab:arr} can be used to obtain the frequency factor using \added{the Einstein relation for diffusion in two dimension \citep{Tsong2005}:}
\begin{equation}
D_0=\nu a^2/4
\label{eq:d_nu}
\end{equation}
where $a=0.3$ nm is the distance between two nearest adsorption sites on ASW assumed in our analysis. This is the expression pertinent to tracer diffusion. The resulting frequency $\nu$ can be used in astrochemical modeling to describe the diffusion of molecules on the surface of ASW. The values of $\nu$ are mostly in the range of $10^8$ to $10^9$ s$^{-1}$. This is orders of magnitude lower than the typical value of the single particle vibrational frequency \added{$\nu_0$}($10^{12}$ to $10^{13}$ s$^{-1}$) \added{that is assumed in models}. If we assume that the pre-exponent $\nu$  for CO$_2$ is $10^8$ s$^{-1}$ instead of $10^{12}$ s$^{-1}$, the $E_{\rm dif}$/$E_{\rm des}$ ratio would be lowered to the 0.6--0.7 range, closer to the ratio for more volatile molecules. Further laboratory study are necessary to quantify the diffusion parameters of less volatile molecules.

\begin{table}[tbp]
\centering
\caption{Activation energy for diffusion $E_{\rm dif}$, binding energies $E_{\rm des}$, their ratio $E_{\rm dif}$/$E_{\rm des}$ , and frequency $\nu$ calculated using Eq~\ref{eq:d_nu} (a - this work, b - \citet{He2016}). \added{The $E_{\rm des}$ distributions are obtained from direct inversion of TPD spectra measured on np-ASW. The lower and higher values of $E_{\rm des}$ corresponds to the desorption energy of multilayer coverage and very low coverage, respectively.} }
\label{tab:ratios}
\begin{tabular}{lcccc}
    &  $E_{\rm dif}$ / K  & $E_{\rm des}$/K & $E_{\rm dif}$/$E_{\rm des}$  & $\nu$ / s$^{-1}$ \\ \hline
CO    & $490\pm12^{a}$ & 870--1600$^{b}$ & 0.31--0.56  & $1.5\times 10^{9}$  \\
N$_2$  & 447$\pm$9$^{a}$ & 790--1320$^{b}$ & 0.34--0.57  & $5.2\times 10^{8}$   \\
O$_2$    & 446$\pm$12$^{a}$ &920--1520$^{b}$ & 0.29--0.48  & $4.5\times 10^{8}$  \\
Ar  & 443$\pm$10$^{a}$ &  930--1420$^{a}$& 0.31--0.48    & $5.3\times 10^{7}$    \\
CH$_4$  & 547$\pm$10$^{a}$ &1100--1600$^{b}$ & 0.34--0.50 & $2.6\times 10^{9}$ \\
H$_2$   &  $<$120$^{a}$  & --  &  --  &  -- \\
D$_2$   &  $<$120$^{a}$  & --  &  --  &  --
\end{tabular}
\end{table}

In conclusion, we presented the first comprehensive set of diffusion parameters (pre-exponent $D_0$ and activation energy $E_{\rm dif}$ in the Arrhenius expression for the diffusion coefficient) for simple molecules diffusing on the pore surface of ASW. Such parameters can be readily applied in the simulation of the evolution of ices in ISM environments.

\acknowledgements
We thank Francis Toriello for technical
assistance. This research was supported by NSF Astronomy \&
Astrophysics Research Grant \#1615897.

\end{document}